\documentclass{article}      

\usepackage{epsfig}
\usepackage{tabularx}
\usepackage{enumitem}
\usepackage{todonotes}
\usepackage{cite}
\usepackage[breaklinks]{hyperref}
\usepackage{breakurl}

\title{Ride Sharing \& Data Privacy: An Analysis of the State of Practice}

\author{Carsten Hesselmann\textsuperscript{1}, Jan Gertheiss\textsuperscript{2}, Jörg P.~Müller\textsuperscript{1}}

\date{}

\begin{document}
\def\UrlBreaks{\do\/\do-}

\maketitle

\begin{table}[h]
\begin{tabular}{cc}
    \footnotemark[1]Clausthal University of Technology, &
    \footnotemark[2]Helmut Schmidt University, \\
    
    Institute of Informatics, &
    School of Economics and \\
    
    38678 Clausthal-Zellerfeld, &
    Social Sciences, \\
    
    Germany & 22043 Hamburg, Germany \\
    
\end{tabular}
\end{table}


\section*{Abstract}

Digital services like ride sharing rely heavily on personal data as individuals have to disclose personal information in order to gain access to the market and exchange their information with other participants; yet, the service provider usually gives little to no information regarding the privacy status of the disclosed information though privacy concerns are a decisive factor for individuals to (not) use these services. We analyzed how popular ride sharing services handle user privacy to assess the current state of practice. The results show that services include a varying set of personal data and offer limited privacy-related features.

\section{Introduction}
\label{sec:introduction}

Privacy is important in private, public and commercial context, which has strongly and repeatedly been emphasized in recent years, e.g. when discussing advancements in artificial intelligence \cite{Brundage.2018,HLEG.2019}, covering responsibility of governments and companies in the digital age \cite{WWWF.2019}, and debating ethics of information collection \cite{Dat.2019,IEEE.2017}. 
By the same token, privacy is relevant when an individual interacts with other people whether it be within or beyond their social circles \cite{Page.2012,Squicciarini.2014,Jagadish.2020}. Communicating and meeting with an unknown person is a common interaction in the sharing economy - a term for (digital) peer-to-peer marketplaces where individuals share their private property with others prior to knowing them. Individuals have varying motivations for sharing \cite{Schor.2016,Milanova.2017} and there are numerous marketplaces which offer services for sharing, for example, apartment rentals and shared rides. 




In order to take part in ride sharing, users typically need to register an account to gain access to the service's website which serves as a peer-to-peer marketplace. The initial communication between users is commonly accompanied by an exchange of personal information which is essential for trust-building \cite{Proserpio.2018}; yet, services that offer a marketplace to share goods usually give little to no information about the privacy of their personal data when asking them to disclose it. In that context, it is important not only whether an individual discloses information to the service's provider but also whether other registered users can access this information since the interaction with others is arguably the main reason for a user to use such a service.

Moreover, the perception of privacy (concerns) and the resulting behavior depend on each individual \cite{Beke.2018}. For that reason, it is relevant for the user to know whether or not any disclosed information is exposed to other registered members of the same marketplace. This is especially important, considering that privacy concerns are a decisive factor for individuals to (not) participate in the sharing economy \cite{Ranzini.2017} 
 and unclear data practices are a critical flaw for any digital marketplace.

As a first step to address this issue we conducted a pilot study and present the results in this paper. We analyze twelve popular ride sharing services in Germany, Austria, and Switzerland, according to various ratings \cite{EntegaPlusGmbH.2014,UtopiaGmbH.2016,VCSVerkehrsClubderSchweiz.2019,VGLVerlagsgesellschaft.2016}. Our goal is to answer the following research questions:

\begin{itemize}
    \item RQ1: What personal data are commonly included in ride sharing services?
    \item RQ2: What privacy settings are available as part of ride sharing services (e.g. change exposure of certain information to other users)?
    \item RQ3: What privacy-related features do services for ride sharing offer (e.g. validation of authenticity)?
\end{itemize}

\section{Analysis of Ride Sharing Services}
\label{sec:serviceanalysis}

For the analysis of ride sharing services, we choose the most popular services in Germany, Austria and Switzerland. We analyze the websites on which the services are offered, and investigate how user data are acquired on these websites. This analysis was limited to those areas and features of the websites,  which are accessible to users including the privacy policies.

Furthermore, the analysis only includes information that is directly linked to the individual, their preferences, or information about their vehicle. Consequently, information relating solely to a ride offer, e.g. price, or locations and routes is not included in this analysis as they are an extensive research area on their own \cite{Schilit.2003,Tsai.2009,Shin.2012,Yin.2018}.

The results cover twelve ride sharing services and 41 data attributes in total (after merging similar attributes, as summarized in Table~\ref{tab:datamerges}). Our analysis shows which data attributes are part of the service and are exposed to other users, as discussed in Section~\ref{subsec:collectionandexposure}. In addition to that, available privacy settings and the validation of information are included as well, as reviewed in Section~\ref{subsec:rsservicefeatures}. This analysis was limited to shared rides. Thus, if cargo transport or the like was offered, it was not included.

\begin{table}[!t]
    \centering
    \caption{Merged data attributes}
    \label{tab:datamerges}
    \begin{tabular}{m{0.255\textwidth} | m{0.675\textwidth}} 
    	\textbf{merged attribute} & \textbf{original attributes} \\
    	\hline
    	contact information & cellphone number, landline number, fax number \\
    	\hline
    	social media & Facebook, Youtube, personal website \\
    	\hline
    	personal description & personal characteristics, self description, things I like \\
    	\hline
    	interests & sport, hobby, movie \\
    	\hline
    	job & job description, job industry \\
    	\hline
    	address & country, city, zip code, street \\
    \end{tabular}
\end{table}

The following steps were carried out to analyze a ride sharing service: (i) register an account (if possible), (ii) complete the profile with data attributes, (iii) check for available profile settings, (iv) review profile pages, (v) create ride offers, (vi) review ride offers, and (vii) book ride offers. 

\subsection{Collection and Exposure of Personal Information}
\label{subsec:collectionandexposure}

Firstly, the analysis includes the attributes of personal information which are part of the ride sharing service, with the collection of this information being either \textit{mandatory} or \textit{optional}. Secondly, the exposure of each attribute (towards other users) is included. A data attribute is (not) exposed on either the \textit{profile page} and/or on the \textit{ride offer}.

\begin{figure}[!t]
	\centering
	\includegraphics[width=1\linewidth]{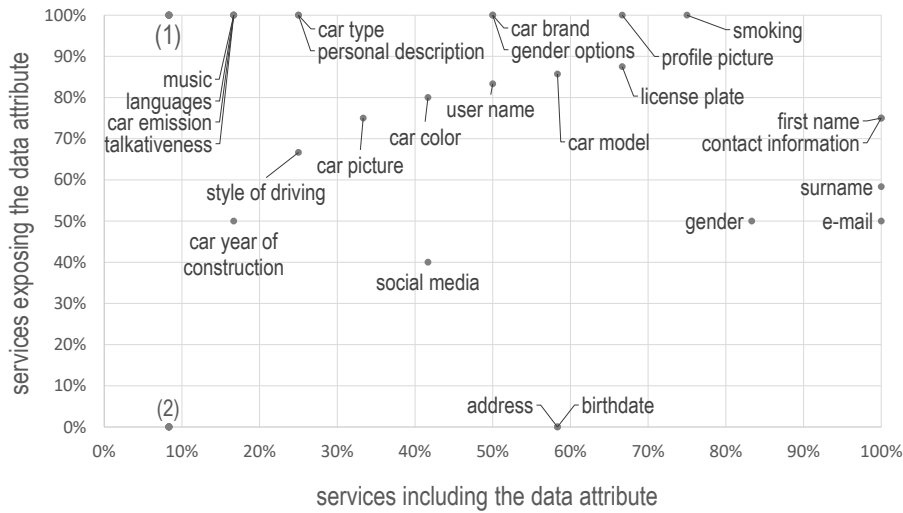}
	\caption{Services that include (x-axis) and expose (y-axis) data attributes. Abbreviations: (1) fuel type, handicapped accessible, interests, job, membership automobile club, phone owner, phone provider, pick-up / drop-off (2) air-condition, bank account, car mileage, country of car registration, driver‘s license, fuel consumption, marital status, PayPal \textit{[date of access: 2021-01-26]}}
	\label{fig:graph-collection-exposure}
\end{figure}

Across all data attributes, the average rate of collection (i.e. how many services include this information) is 35\%. This shows that the set of data attributes across ride sharing services is not standardized. The average type of collection across all 41 data attributes are 29\% mandatory and 71\% optional, which advocates a tendency towards a user-friendly type of collection though individuals tend to over-disclose optional information \cite{Preibusch.2013}. Certain information is exposed in a reduced fashion; for instance, if the date of birth was disclosed, only the age (in years) was made accessible to other users. Ten data attributes were never displayed for other users, which makes the collection of this information questionable from a user perspective. Fig.~\ref{fig:graph-collection-exposure} displays the full set of included data attributes from all analyzed services. 

The data points are widely spread, indicating the diverse inclusion of personal information and a lack of standardization. In addition, our results show a great variety among the analyzed services in terms of the included user data; the set of collected data attributes ranges between 5 and 30 attributes while the set of exposed data attributes varies between 4 and 18 attributes -- the statistics for all analyzed ride sharing services are provided in Table~\ref{tab:statisticservices}.

\begin{table}[!b]
    \centering
    \caption{Analysis results: (1) bessermitfahren.de, (2) blablacar.de, (3) clickapoint.com, (4) e-carpooling.ch, (5) fahrgemeinschaft.de, (6) foahstmit.at, (7) greendrive.at, (8) mifaz.de, (9) mitfahrangebot.de, (10) mitfahrportal.de, (11) pendlerportal.de, (12) twogo.com \textit{[date of access: 2021-01-26]}}
    \label{tab:statisticservices}
    \begin{tabular}{m{0.175\textwidth} 
    m{0.03\textwidth} m{0.03\textwidth} m{0.03\textwidth}
    m{0.03\textwidth} m{0.03\textwidth} m{0.03\textwidth}
    m{0.03\textwidth} m{0.03\textwidth} m{0.03\textwidth}
    m{0.03\textwidth} m{0.03\textwidth} m{0.03\textwidth}}
    	& (1) & (2) & (3) & (4) & (5) & (6) & (7) & (8) & (9) & (10) & (11) & (12) \\
    	\hline
    	\hline
    	collected data & 9 & 23 & 13 & 16 & 14 & 5 & 6 & 16 & 11 & 30 & 15 & 13 \\
    	\hline
    	exposed data & 9 & 11 & 11 & 11 & 11 & 5 & 4 & 15 & 5 & 18 & 8 & 10 \\
    	\hline
    	\hline
    	mandatory & 3 & 14 & 3 & 9 & 4 & 4 & 3 & 2 & 8 & 4 & 8 & 9 \\
    	\hline
    	optional & 6 & 9 & 10 & 7 & 10 & 1 & 3 & 14 & 3 & 26 & 7 & 4\\
    	\hline
    	\hline
    	profile page & 0 & 2 & 0 & 1 & 3 & 0 & 0 & 0 & 0 & 5 & 0 & 0 \\
    	\hline
    	ride offer & 9 & 4 & 6 & 5 & 3 & 5 & 4 & 7 & 0 & 11 & 8 & 10 \\
    	\hline
    	both & 0 & 5 & 5 & 5 & 5 & 0 & 0 & 8 & 5 & 2 & 0 & 0 \\
    \end{tabular}
\end{table}


Furthermore, relating these numbers shows noteworthy results. For four ride services the percentage of exposed, i.e. visible for other users, data attributes is close to or below 50\% (e.g. ride services blablacar.de and pendlerportal.de). This raises the question why the user should disclose their personal information to the service provider if not even half of it is accessible to the other users -- especially in the context of sharing rides, where by design the interaction with other users is arguably the main reason for an individual to use such a service. This question is further aggravated by the fact that explanations to the user on why this information should be disclosed are lacking in almost all instances as the privacy policies provided basic legal terminology. Only a limited number of services mention available privacy settings (e.g. change the expose of information towards other users) in their privacy policies.

Moreover, some services distinguish between the user's ride offers and profile by separating them on different pages. The specific focus on the profile page differs as some services show exactly the same information on the profile page as they do with the ride offer. However, four services provide a dedicated profile page, promoting the user's identification with their profile.

\subsection{Privacy-Related Features}
\label{subsec:rsservicefeatures}

Seven services allow the user to express individual privacy preferences for some data attributes (Table~\ref{tab:privacysettings} provides an overview). In most cases, the availability of profile settings is not communicated to the user at the time of first disclosing the information. In addition to that, the validation of information is compared in detail. Our analysis shows that five ride sharing services offer validation of the authenticity of certain personal information, e.g.~via legal documents or means of communication (as shown in Table~\ref{tab:confirmedinformation}).

\begin{table}[!b]
    \centering
    \caption{Privacy setting [date of access: 2021-01-26]}
    \label{tab:privacysettings}
    \begin{tabular}{m{0.25\textwidth} | m{0.68\textwidth}} 
    \textbf{service provider} & \textbf{privacy settings} \\
    \hline
    bessermitfahren.de & It is mandatory for the user to show the \textit{e-mail} or the \textit{phone number} in the ride offer. \\
    \hline
    blablacar.de & The \textit{phone number} can be shown with the ride offer or made visible only for users who booked the ride. \\
    \hline
    clickapoint.com & The user can decide whether their information for direct communication is made accessible for others or not. If not, the only way to contact the user is to use the internal messaging system. \\
    \hline
    fahrgemeinschaft.de & When creating a ride offer the user can choose which information is shown to all users, shown only to other members of the service, shown only on request, or not shown at all. This can be set for \textit{surname}, \textit{e-mail}, \textit{phone number}, \textit{license plate}, and \textit{car model}. \\
    \hline
    greendrive.at & The user can decide whether \textit{e-mail} is visible for other users before or after booking a ride. \\
    \hline
    mifaz.de & The user can decide to (not) show \textit{first name}, \textit{surname}, and \textit{phone number}. \\
    \hline
    mitfahrportal.de & When creating a ride offer the user can choose whether \textit{e-mail} and/or \textit{phone number} is to be included in the ride offer. It is mandatory to choose one of the two. The user can choose whether the \textit{phone number} is shown to all users or only to registered members. \\
    \end{tabular}
\end{table}

\begin{table}[!t]
    \centering
    \caption{Information validation [date of access: 2021-01-26]}
    \label{tab:confirmedinformation}
    \begin{tabular}{m{0.25\textwidth} | m{0.68\textwidth}} 
    \textbf{service provider} & \textbf{information validation} \\
    \hline
    blablacar.de & \textit{E-mail} and \textit{phone number} can be validated. \\
    \hline
    clickapoint.com & The \textit{e-mail} is validated. \\
    \hline
    e-carpooling.ch & The user profile can be validated, it is necessary for the user to upload a copy of their driving license/ID. \\
    \hline
    fahrgemeinschaft.de & \textit{Automobile club membership} and \textit{e-mail} are validated. \\
    \hline
    twogo.com & The service offers to validate the \textit{phone number}. \\
    \end{tabular}
\end{table}

\subsection{Analysis Results}
The amount of personal information that is included in shared mobility services varies greatly. The largest set of data attributes included in one service was 30 data attributes and the smallest set included 5 data attributes. This outlines the heterogeneity of offered services and indicates a lack of standardization with respect to the personal information users provide for using shared mobility services. Moreover, the fact that our analysis includes 41 different data attributes (after merging similar attributes) that users are asked to disclose stresses the lack of standardization once more. These results answer RQ1.

To answer RQ2, we investigated whether disclosed personal information is accessible for other users. On average, services show 73\% of the disclosed personal information to other users. This means one quarter of the disclosed personal information is not part of any user-to-user interaction and remains only with the service provider. Some services are close to or even below the 50\% margin, which emphasizes how different the exposure of personal information towards other users can be depending on the choice of service. How this affects user satisfaction remains a topic for further research.

As for RQ3, we found that only a limited number of shared mobility services offer privacy-related features. Moreover, the range and type of these features varies greatly. Some services offer or require the user to complete a process to validate the authenticity of personal information while other services offer privacy settings to change the exposure of certain data. Those can affect the communication with other users, matchmaking among groups, and whether specific information (e.g. e-mail address or phone number) is accessible for others.

\section{Discussion and Outlook}
\label{sec:discussion}

Our findings show that inclusion of personal data varies greatly between ride sharing services. This includes the user's personal data and how user-to-user privacy is handled, i.e. how much of the personal information an individual discloses is exposed to other individuals. Most services only expose a limited set of data attributes without giving explanations for it, which is debatable considering the nature of a peer-to-peer marketplace. How these differences in user-to-user privacy affect, for example, the user's satisfaction with the service or the perceived usability, could be extended on.

Moreover, the variety of included data attributes and also the number of included attributes clearly indicate a lack of standardization among ride sharing services. Having insights into the data processing beyond the user's scope could complement our results, as this analysis was limited to parts which are accessible for users. With the study presented in this paper we have laid the foundation for the next steps of our research. We plan to conduct an online survey in which participants are asked whether the in this paper presented privacy-related differences among ride sharing services would affect the of the participants' behavior when using ride sharing services, for example, when decide to disclose some of their personal information.


\bibliographystyle{alpha}
\newcommand{\etalchar}[1]{$^{#1}$}

\end{document}